\documentclass[letterpaper]{article} 
\usepackage{aaai2026}  
\usepackage{times}  
\usepackage{helvet}  
\usepackage{courier}  
\usepackage[hyphens]{url}  
\usepackage{graphicx} 
\usepackage{natbib}  
\usepackage{caption} 
\usepackage{booktabs}
\usepackage{amsmath}
\usepackage{algorithm}
\usepackage{algorithmic}
\usepackage{makecell}
\usepackage{tabularx}
\usepackage{booktabs} 
\usepackage{graphicx}
\usepackage{soul}
\usepackage{xcolor}
\usepackage{algorithm}
\usepackage{algorithmic}

\usepackage{newfloat}
\usepackage{listings}
\DeclareCaptionStyle{ruled}{labelfont=normalfont,labelsep=colon,strut=off} 
\floatstyle{ruled}
\newfloat{listing}{tb}{lst}{}
\floatname{listing}{Listing}
\title{Improving Q-Learning for Real-World Control: A Case Study in Series Hybrid Agricultural Tractors}
\author{
    Hend Abououf,
    Sidra Ghayour Bhatti,
    Qadeer Ahmed
}
\affiliations{
    \textsuperscript{\rm }Mechanical and Aerospace Engineering, Center for Automotive Research\\

    The Ohio State University, Ohio, USA\\
    abououf.1@osu.edu, bhatti.39@osu.edu, ahmed.358@osu.edu
}

\usepackage{bibentry}
\nocopyright
\begin{document}

\maketitle

\begin{abstract}
The variable and unpredictable load demands in hybrid agricultural tractors make it difficult to design optimal rule-based energy management strategies, motivating the use of adaptive, learning-based control. However, existing approaches often rely on basic fuel-based rewards and do not leverage expert demonstrations to accelerate training. In this paper, first, the performance of Q-value-based reinforcement learning algorithms is evaluated for powertrain control in a hybrid agricultural tractor. Three algorithms, Double Q-Learning (DQL), Deep Q-Networks (DQN), and Double DQN (DDQN), are compared in terms of convergence speed and policy optimality. Second, a piecewise domain-specific reward-shaping strategy is introduced to improve learning efficiency and steer agent behavior toward engine fuel-efficient operating regions. Third, the design of the experience replay buffer is examined, with a focus on the effects of seeding the buffer with expert demonstrations and analyzing how different types of expert policies influence convergence dynamics and final performance. Experimental results demonstrate that (1) DDQN achieves 70\% faster convergence than DQN in this application domain, (2) the proposed reward shaping method effectively biases the learned policy toward fuel-efficient outcomes, and (3) initializing the replay buffer with structured expert data leads to a 33\% improvement in convergence speed.
\end{abstract}


\section{Introduction}

Agricultural tractors operate under various duty cycles, each characterized by different power demands.  For example, the power requirement for the same chisel plow operation can change from 100 kW to 320 kW based on the depth of the equipment. Whereas the 34 inch rotary Hoe requires 5.5 kW.  As a result, when designing a hybrid tractor, fuel consumption varies significantly depending on the duty cycle. Using a rule-based controller, Table \ref{tab:results} demonstrates that certain duty cycles result in higher fuel consumption compared to conventional tractors. However, by adjusting the rule-based control strategy, fuel efficiency improvements were observed. Given this variability, it is essential to develop a controller that can dynamically adapt to varying tractor loads and optimize control strategies accordingly.

\begin{table}[t]
\small 
\centering
\caption{Fuel and SOC Results for 13-Minute Planter Operation at 216 kW Load}
\label{tab:results}
\begin{tabular}{lccc}
\toprule
Metric & Conventional & Control 1 & Control 2 \\
\midrule
Fuel Consumption (gal) & 1.69 & 1.53 & 1.20 \\
Fuel Reduction (\%)     & --   & -9.4\% & -29\% \\
SOC at End (\%)         & --   & 71\%   & 65\% \\
\bottomrule
\end{tabular}
\end{table}


In 2014, reinforcement learning (RL) was first applied to power management in hybrid on-road vehicles \cite{b1}. The authors employed Temporal Difference Learning to optimize energy distribution between the internal combustion engine (ICE) and the electric motor (EM) in hybrid electric vehicles (HEVs). Their model-free approach effectively reduced fuel consumption by 42\% compared to rule-based control methods and adapted well to real-time stochastic driving environments. However, the approach incurred high computational costs. Building on this concept, several studies have focused on optimizing engine performance. Ahmadian et al. \cite{ahmadian2023} proposed a Q-learning-based energy management strategy for series-parallel HEVs to improve fuel efficiency and battery life without prior knowledge of the driving cycle. The agent learns the optimal power split based on battery state-of-charge (SOC), power demand, and vehicle speed, with a reward function balancing fuel savings and battery health.The results showed up to 65\% battery life improvement and a 1.25\% reduction in fuel consumption compared to rule-based strategies. Similarly, Mousa \cite{mousa2023} introduced an Extended Deep Q-Network (E-DQN) for energy management in plug-in HEVs, enhancing traditional rule-based and heuristic strategies. The E-DQN achieved 98.47\% of the Dynamic Programming (DP) benchmark. Another study by Han et al. \cite{Han2019} applied Double Deep Q-Learning (DDQL) for energy management in hybrid electric tracked vehicles (HETVs), addressing overestimation in conventional Deep Q-Learning. DDQL achieved 93.2\% of DP performance and 7.1\% improvement over DQL, while maintaining SOC stability and offering faster convergence and better generalization.

These studies were conducted on passenger vehicles with relatively stable duty cycles. Zhang et al. \cite{Zhang2021} applied DDQN under variable driving conditions, shaping the reward based on SOC, achieving 92.7\% of DP results. DDQN was also implemented on a hybrid agricultural tractor \cite{Zhang_2025}, achieving a 4.5\% reduction in fuel consumption over a power-following strategy. The reward functions in these studies were based on fuel use and SOC. Despite these results, engine efficiency remains overlooked. Operating at low power reduces fuel use but can yield efficiencies as low as 20\%.

To address this, the current study introduces a piecewise reward shaping function based on engine efficiency. Furthermore, DDQN’s slow convergence can be improved by preseeding the replay buffer with expert offline data. Hester et al. \cite{Hester2017} showed that combining TD updates with supervised large-margin classification loss and prioritized replay significantly accelerates learning, outperforming the best demonstrator in 14 of 42 Atari games. This paper aims to improve policy optimality and convergence speed by integrating engine efficiency-based reward shaping with expert data preseeding for hybrid agricultural tractor energy management.


The primary contributions of this study are summarized as follows:
(1) Implementation of Tabular Reinforcement Learning: Explored the feasibility of applying tabular Q-Learning and Double Q-Learning (DQL) to hybrid vehicle energy management, emphasizing the limitations imposed by the large state-action space.
(2) Application of Deep Reinforcement Learning: Developed and evaluated Deep Q-Network (DQN) and Double Deep Q-Network (DDQN) models to address scalability challenges inherent in tabular methods.
(3) Development of a Piecewise Reward Shaping Function: Improved learning optimality by designing a reward shaping function that discourages operation in low-efficiency engine regions and promotes operation at peak engine efficiency.
(4) Expert Knowledge Integration via Seeding: Enhanced learning efficiency by initializing the replay buffer with expert-generated experiences from Dynamic Programming (DP) and rule-based controllers, and investigated their impact on convergence and performance.
(5) Evaluation of Seeding Impact: Conducted a comparative analysis between seeded and non-seeded DDQN models, using various expert datasets to quantify the influence of replay buffer initialization on training efficiency and control effectiveness.
(6) Benchmarking Against Dynamic Programming (DP): Benchmarked the performance of learning-based approaches against an optimal DP solution to evaluate relative effectiveness in reducing fuel consumption.

\section{Model Framework}
\begin{figure}
    \centering
    \includegraphics[width=\linewidth]{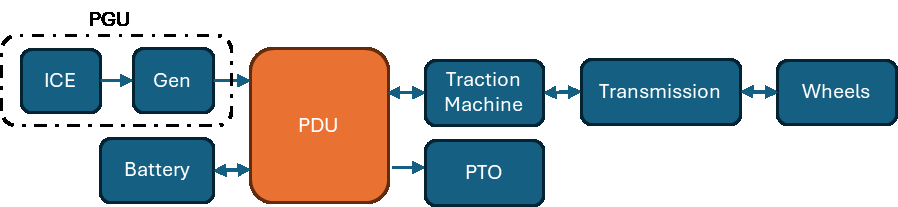} 
    \caption{Schematic of the series hybrid powertrain integrating a power generation unit, battery, and traction motor. PDU is the power distribution unit}
    \label{fig:re_powertrain} 
\end{figure}
The powertrain architecture, illustrated in Figure~\ref{fig:re_powertrain}, follows a series hybrid configuration. The primary objective is to determine the optimal power split between the Power Generation Unit (PGU) and the battery, in order to minimize fuel consumption. The system dynamics are governed by the energy balance equation:

\begin{equation}
    P_{\text{dem}} = P_{\text{batt}} + P_{\text{ICE}},
    \label{eq:energy_balance}
\end{equation}

\noindent where \( P_{\text{dem}} \) is the total power demand, composed of battery power \( P_{\text{batt}} \) and internal combustion engine (ICE) power \( P_{\text{ICE}} \).

The state space is defined by the discretized power request \( P_{\text{dem}} \) and the battery state of charge (SOC):

\begin{equation}
    S = [P_{\text{dem}}, \text{SOC}].
\end{equation}

\noindent SOC is constrained within operational limits:

\begin{equation}
    \text{SOC}_{\text{min}} \leq \text{SOC} \leq \text{SOC}_{\text{max}}.
\end{equation}

The action space is represented by the discretized battery power:

\begin{equation}
    a = [P_{\text{batt}}].
\end{equation}

\noindent Battery power is bounded by both physical and operational constraints:

\begin{equation}
    \text{lb} = \max \left[ \left( \frac{(\text{SOC}_{\text{max}} - \text{SOC}(j)) Q_{\text{batt}} V_{\text{oc}}}{-t_s} \right),\ -P_{\text{b,max}},\ \Delta P \right]
\end{equation}

\begin{equation}
    \text{ub} = \min \left[ \left( \frac{(\text{SOC}_{\text{min}} - \text{SOC}(j)) Q_{\text{batt}} V_{\text{oc}}}{-t_s} \right),\ P_{\text{b,max}},\ P_{\text{dem}}(j) \right]
\end{equation}

\noindent where \( \Delta P = P_{e,\text{max}} - P_{\text{dem}}(j) \), \( Q_{\text{batt}} \) is the nominal battery capacity, \( V_{\text{oc}} \) is the open-circuit voltage, and \( t_s \) is the sampling time.

The feasible action space is therefore given by:

\begin{equation}
    \text{lb} \leq P_{\text{batt}}(j) \leq \text{ub}.
\end{equation}

The initial reward function is defined as:

\begin{equation}
    r = -\dot{m} \Delta t
\end{equation}

\noindent where \( \dot{m} \) is the instantaneous fuel consumption rate, $\Delta t$ is the step size.

\section{Double Q-Learning}

Double Q-learning is an extension of the standard Q-learning algorithm that addresses the overestimation bias introduced by the maximization step in the target value computation. In standard Q-learning, the same value function is used to both select and evaluate the next action, which often results in overly optimistic value estimates.

To mitigate this, Double Q-learning maintains two independent action-value functions, \( Q_A \) and \( Q_B \), and decouples the action selection from evaluation. During training, one of the two estimators is randomly selected for updating. For instance, if \( Q_A \) is chosen for update, the action is selected using \( Q_A \), but evaluated using \( Q_B \). The target value is computed as:

\begin{align}
    &y_t = r_t + \gamma Q_B\left(s_{t+1}, \arg\max_a Q_A(s_{t+1}, a) \right) \\
    &Q_A(s_t, a_t) \leftarrow Q_A(s_t, a_t) + \alpha \left[ y_t - Q_A(s_t, a_t) \right]
\end{align}

This formulation reduces the overestimation of action values, resulting in more stable and reliable policy learning, especially in environments with high uncertainty or noise \cite{b16}, \cite{b17}.
\subsection{\textbf{Results of Double Q-Learning}}

\begin{figure}
    \centering
    \includegraphics[width=\linewidth]{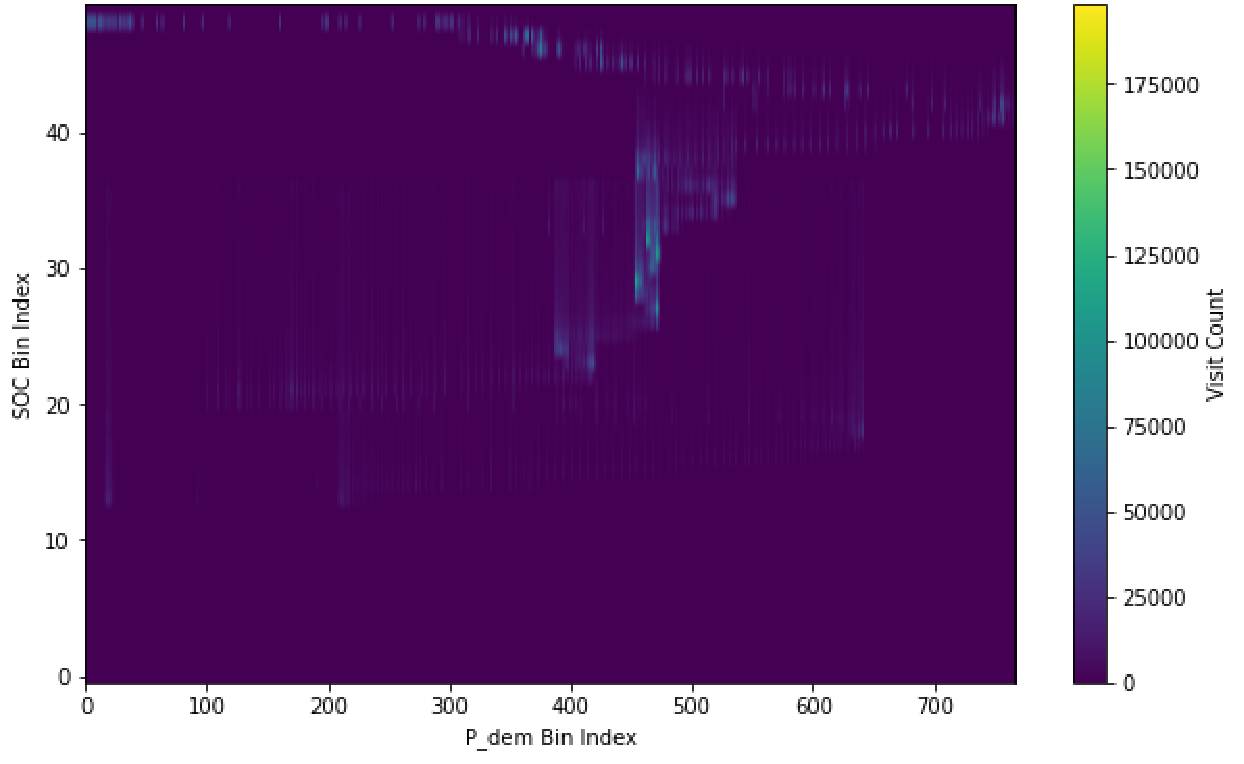}
    \caption{State Visit heatmap}
    \label{fig:heatmap}
\end{figure}

Since Double Q-learning is known for its improved stability over standard Q-learning, it was initially implemented in this study. However, the results demonstrate that tabular Q-learning approaches are not suitable for the current application for several reasons:

\begin{enumerate}
    \item \textbf{Extremely Large Environment Space:} The discretized environment comprises 50 bins for the state-of-charge (SOC), 766 bins for the required power, and 1600 bins for the action space. This results in over 61 million entries in the Q-table, far exceeding the practical limits of tabular Q-learning, which is typically suitable only for small state-action spaces~\cite{b18}. As a result, the agent remains significantly undertrained, with the majority of state-action pairs rarely or never encountered even after 100,000 episodes. Figure~\ref{fig:heatmap} shows the visit frequency for each state pair $(P_{dem}, \text{SOC})$, highlighting that many state pairs are never visited, and only a few are visited repeatedly.
    
    \item \textbf{Limitations of tabular methods:} Unlike function approximation methods, tabular Q-learning cannot generalize or estimate values between discrete bins. Reducing the number of bins to decrease the environment size is not feasible, as it would significantly degrade the resolution and quality of the control strategy.
    
    \item \textbf{Poor correlation between Q-values and rewards:} To assess whether the Q-values accurately represent the reward structure, the Pearson correlation coefficient was calculated. 
    
    The Pearson correlation coefficient is a measure of the linear relationship between the covariance of Q-values and the reward $COV(Q,r)$. It is represented by the covariance of the two variables divided by the product of their standard deviation $\rho_Q \cdot \rho_r$.
    \begin{equation}
        \rho_{Q,r}=\frac{COV(Q,r)}{\rho_Q \cdot \rho_r}
    \end{equation}

\[
\quad
\rho_{Q,r} \in
\begin{cases}
+1 & \text{Perfect positive correlation} \\
-1 & \text{Perfect negative correlation} \\
[0.5, 1) & \text{Strong positive correlation}  \\
(-1, -0.5] & \text{Strong negative correlation}  \\
[0.3, 0.5) & \text{Moderate positive correlation} \\
(-0.5, -0.3] & \text{Moderate negative correlation} \\
(0, 0.3) & \text{Weak positive correlation} \\
(-0.3, 0) & \text{Weak negative correlation} \\
0 & \text{No correlation}
\end{cases}
\]
    It is found that $\rho_{Q,r}=-0.213$; this weak negative correlation indicates that the learned Q-function does not properly capture the reward signal, thereby failing to guide optimal decision-making.

\end{enumerate}

\begin{figure}
    \centering
    \includegraphics[width=\linewidth]{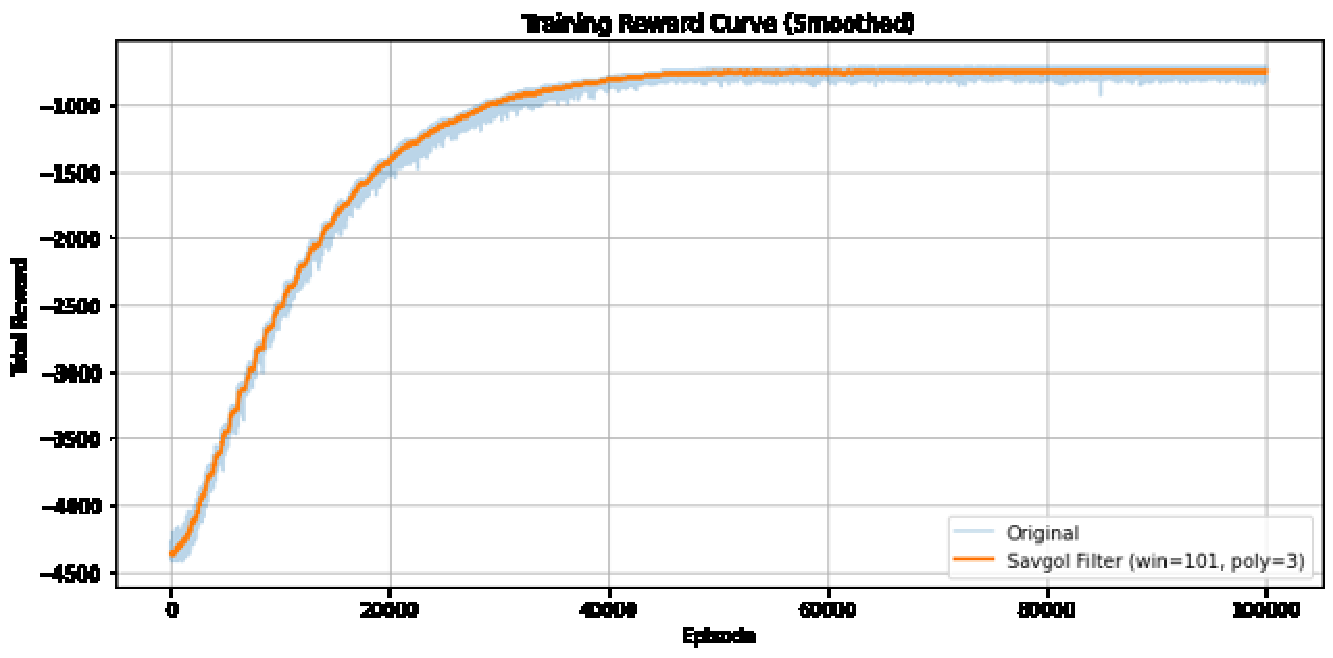}
    \caption{Double Q-learning reward}
    \label{fig:DQL_reward}
\end{figure}

\begin{figure}
    \centering
    \includegraphics[width=\linewidth]{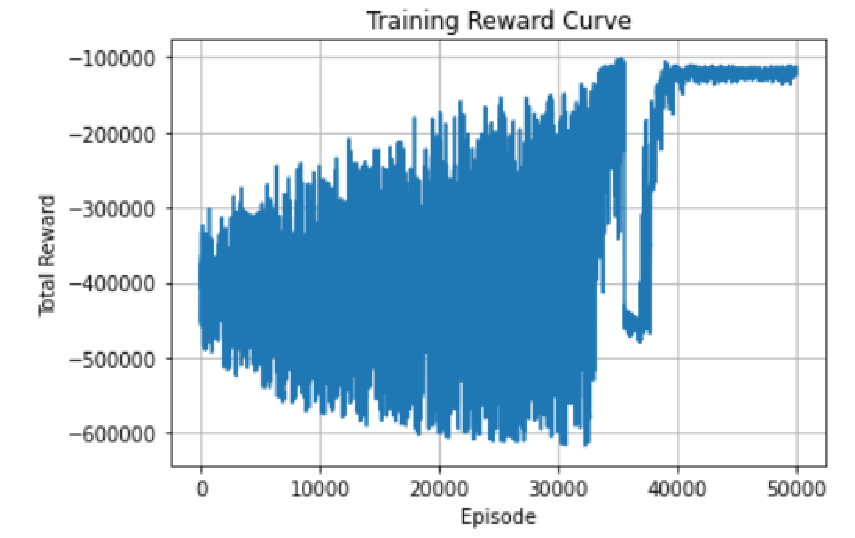}
    \caption{Deep Q-Network reward}
    \label{fig:DQN_reward}
\end{figure}

\section{Deep Q-Network}

\subsection{\textbf{Deep Q-Network}}
The Deep Q-Network (DQN) extends traditional Q-learning by approximating the action-value function $Q(s, a)$ using a neural network, allowing it to scale to high-dimensional state spaces. In DQN, the same network is used for both action selection and evaluation, which can lead to overestimation of Q-values. The target value for training is computed as:

\begin{equation} y_k = r_k + \gamma \cdot \max_{a'} Q(s_{k+1}, a'; \theta^-) \end{equation}

\noindent where $\theta^-$ denotes the parameters of a separate target network that is periodically updated to match the online network parameters. The action-value function is updated using the temporal difference method:

\begin{equation} Q(s_k, a_k) \leftarrow Q(s_k, a_k) + \alpha \left[ y_k - Q(s_k, a_k) \right] \end{equation}

By using a target network and experience replay, DQN stabilizes the training process and has demonstrated success in complex control tasks, although it remains prone to overestimation errors compared to DDQN \cite{b16}.

\subsection{\textbf{Double Deep Q-Network}}
In the Double Deep Q-Network (DDQN) framework, a neural network is employed to estimate the action-value function $Q(s, a)$. To mitigate the overestimation bias observed in DQN, DDQN decouples the action selection and evaluation steps when computing the target value. The target for training is computed as:

\begin{equation}
    y_k = r_k + \gamma \cdot Q\left(s_{k+1}, \arg\max_{a'} Q(s_{k+1}, a'; \theta), \theta^- \right)
\end{equation}

\noindent where $\theta$ represents the parameters of the online network used for action selection, and $\theta^-$ denotes the parameters of the target network used for evaluation. The Q-value is then updated using the temporal difference update:

\begin{equation}
    Q(s_k, a_k) \leftarrow Q(s_k, a_k) + \alpha \left[ y_k - Q(s_k, a_k) \right]
\end{equation}

This decoupling strategy reduces the overestimation bias in Q-learning. DDQN has been shown to improve learning stability and performance in various environments \cite{b16} .

\begin{figure}[h]
    \centering
    \includegraphics[width=\linewidth]{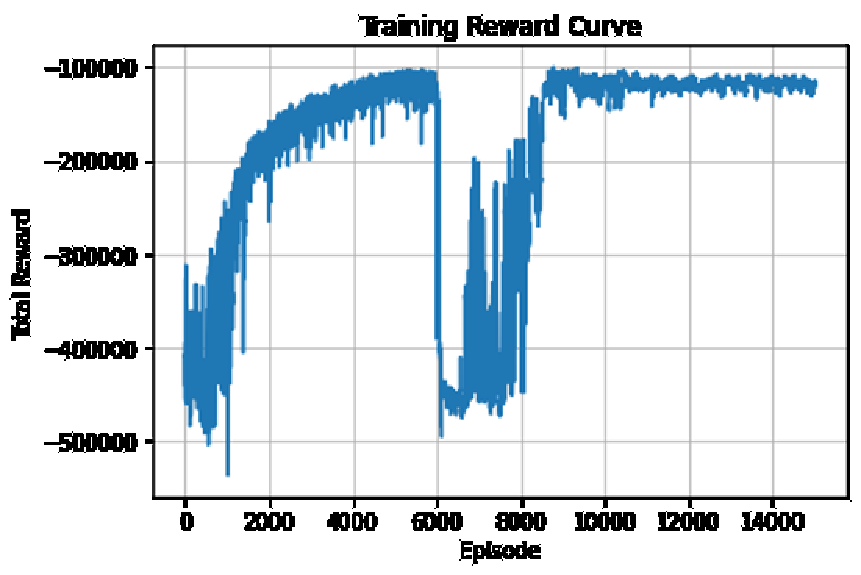}
    \caption{DDQN reward plot}
    \label{fig:DDQN_RB_rewards}
\end{figure}

\begin{figure*}
    \centering
    \includegraphics[width=\linewidth]{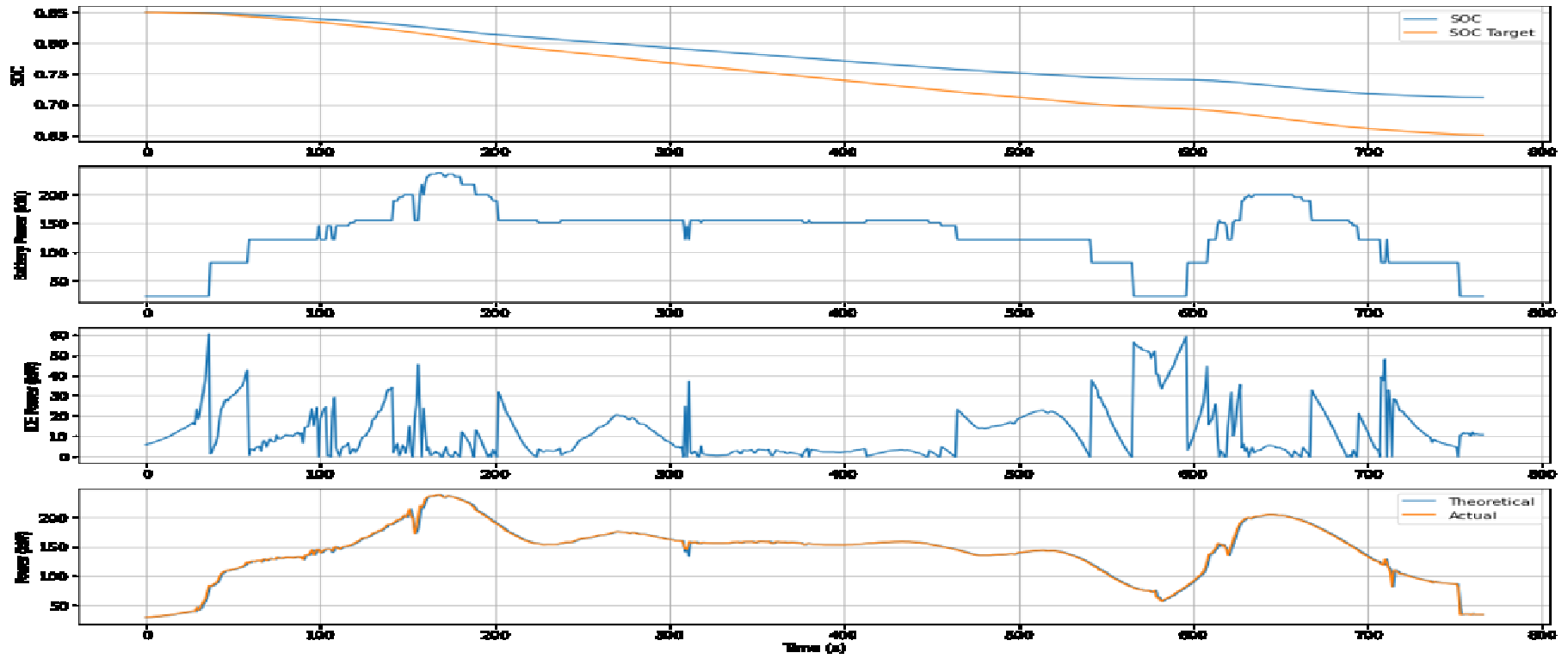}
    \caption{SOC trajectory, Battery power, Engine Power and total power based on DQN and DDQN optimal policy}
    \label{fig:DDQN_RB_power}
\end{figure*}
\subsection{\textbf{Deep Q-Network Result}}
Figure~\ref{fig:DQN_reward} illustrates the convergence behavior of the Deep Q-Network (DQN), which stabilizes after approximately 40,000 episodes. In contrast, the Double Deep Q-Network (DDQN) demonstrates significantly faster convergence, reaching stability after around 9,000 episodes, as shown in Figure~\ref{fig:DDQN_RB_rewards}. Despite this disparity in convergence speed, both methods achieve the same fuel consumption of 0.34 gallons. It is noted that the engine predominantly operates in the region below $60,\text{kW}$ as shown in figure~\ref{fig:DDQN_RB_power} , which corresponds to a low-efficiency regime, resulting in an average thermal efficiency of approximately 20--25\%. 

\section{Reward Shaping}

Since Double Deep Q-Network (DDQN) exhibits significantly faster convergence compared to Deep Q-Network (DQN), it was selected as the final learning algorithm. The state-of-charge (SOC) space is discretized into 200 bins, while the action space consisted of 1600 discrete actions. Initially, the reward function was defined solely in terms of minimizing fuel consumption. Although this approach effectively reduced overall fuel usage, it resulted in suboptimal engine performance, with thermal efficiencies ranging between 20--25\%. This was primarily due to the agent frequently operating the 275kW engine at power levels below, 60kW, an inefficient and generally discouraged operating region.

To overcome this limitation, a piecewise reward shaping function was designed to encourage the agent to select power levels corresponding to higher engine efficiency. Based on analysis of the engine efficiency map, it was observed that the engine achieves maximum efficiency, approximately 42--44\%, when operating between 110kW and 190kW. Furthermore, for low-power demand scenarios, it is more efficient to shut down the engine and rely solely on battery power. Accordingly, the reward function was redefined as follows where $\eta_{eng}$ is the engine efficiency at the current step operating point and $P_{eng}$ is the engine power at the current time step:
\begin{equation*}
\small
r =
\begin{cases}
0.46 + 2, &  P_{\text{eng}} = 0 \\
\eta_{\text{eng}} + 1, &  1.1e^5 \leq P_{\text{eng}} \leq 1.9e^5 \\
-0.58 - 2e^{-6} (1.1e^5 - P_{\text{eng}}) - \eta_{\text{eng}} , &  0 < P_{\text{eng}} < 1.1e^5 \\
-0.06 + 6e^{-6} (1.9e^5 - P_{\text{eng}}) - \eta_{\text{eng}} , &  1.9e^5 < P_{\text{eng}} < 2.75e^5 \\
-2, & \text{otherwise}
\end{cases}
\end{equation*}

\subsection{Reward Shaping Results}
\begin{figure}[h]
    \centering
    \includegraphics[width=\linewidth]{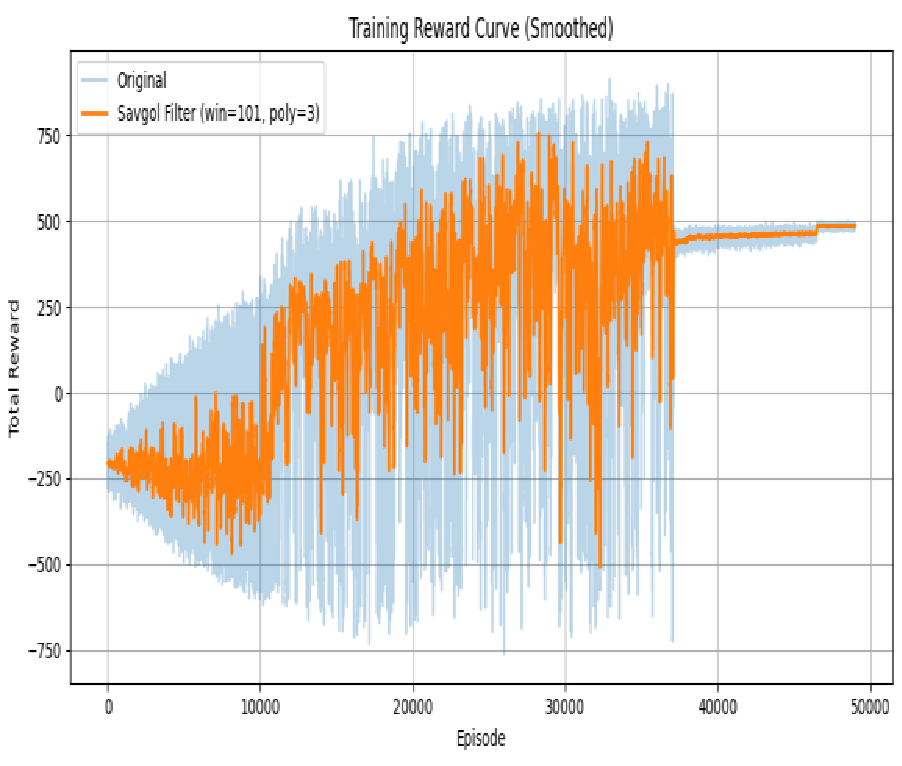}
    \caption{Reward Shaping Reward Plot}
    \label{fig:reward_shaping_reward}
\end{figure}

By integrating the proposed piecewise reward shaping function with learning rate 0.001 and starting with $\epsilon = 1$, the DDQN agent achieved convergence after approximately 37,000 episodes, as illustrated in Figure~\ref{fig:reward_shaping_reward}. Table~\ref{reward_shaping_results} presents a comparison of the DDQN model with reward shaping against the baseline DDQN without reward shaping, dynamic programming (DP) benchmark results, and a conventional vehicle control strategy.
\begin{table}[h]
\centering
\caption{Comparison of Methods Based on Fuel Consumption (FC), Efficiency, and Final SOC}
\small
\begin{tabular}{lccc}
\toprule
\textbf{Method} & \textbf{FC (gal)} & \textbf{Efficiency} & \textbf{Final SOC} \\
\midrule
1.Conventional & 1.95 & 41\% & -- \\
\makecell[l]{2.Baseline DDQN \\ (No Reward Shaping)} & 0.30 & 20--25\% & 70\% \\
\makecell[l]{3.DDQN \\ (Reward Shaping)} & 1.75 & 43\% & 83.6\% \\
4.DP & 1.70 & 43.67\% & 83.6\% \\
\bottomrule
\end{tabular}
\label{reward_shaping_results}
\end{table}

Although the baseline DDQN without reward shaping relies more heavily on battery power, resulting in the lowest fuel consumption of $0.3~\text{gal}$, it operates with very low engine efficiency. In contrast, the DDQN model incorporating reward shaping reduces fuel consumption by 12.8\% compared to the conventional control strategy under the same duty cycle, while also improving engine efficiency by 4.8\%. Moreover, it achieves 97\% of the fuel efficiency and 98.4\% of the thermal efficiency attained by the dynamic programming (DP) benchmark, all while maintaining the same final SOC. These results demonstrate that the proposed piecewise reward shaping function effectively improves both fuel economy and engine efficiency, delivering near-optimal performance that closely approaches the DP benchmark and significantly outperforms the conventional baseline. 

\begin{figure}[h]
    \centering
    \includegraphics[width=\linewidth]{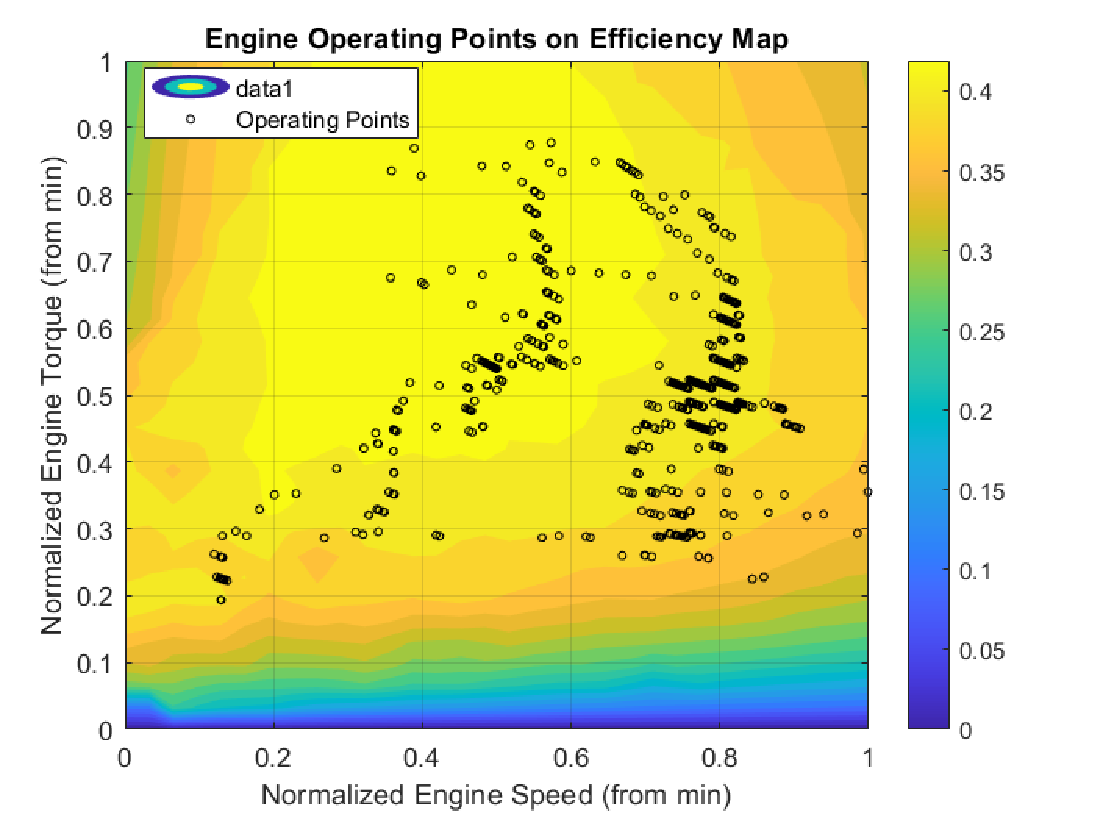}
    \caption{normalized Engine Operating Points for Conventional tractor }
    \label{fig:Conventional_OP}
\end{figure}

\begin{figure}[h]
    \centering
    \includegraphics[width=0.9\linewidth]{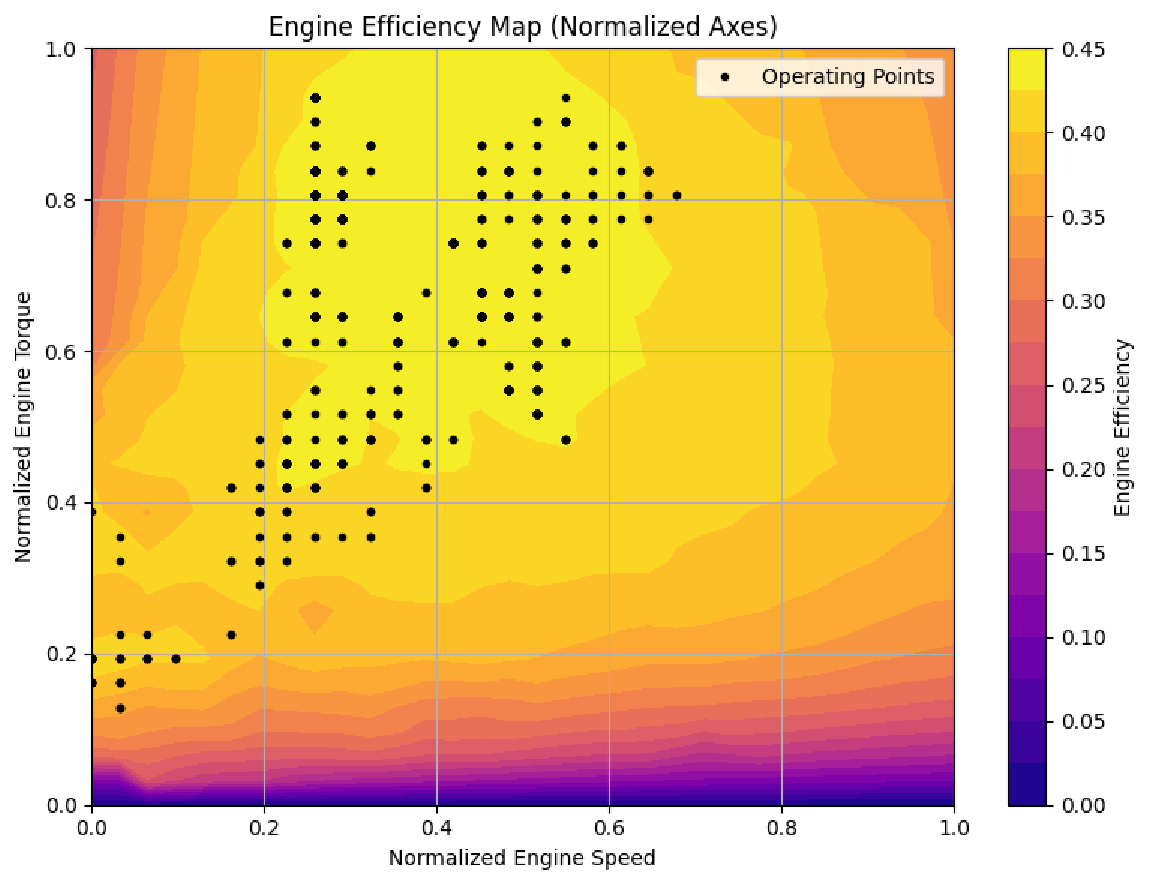}
    \caption{normalized Engine Operating Points for DDQN with reward shaping}
    \label{fig:DDQN_no_seeding_OP}
\end{figure}

Figures~\ref{fig:Conventional_OP} and~\ref{fig:DDQN_no_seeding_OP} show the distribution of engine operating points on the normalized efficiency map for the conventional strategy and DDQN with reward shaping, respectively. In the case of DDQN, the operating points are more concentrated within the high-efficiency region, further validating the effectiveness of the reward shaping approach.

\begin{table*}[h]
\centering
\caption{Comprehensive Comparison of Methods Based on Fuel Consumption (FC), Efficiency, Final SOC, Convergence, and Relative Performance}
\small
\resizebox{\textwidth}{!}{%
\begin{tabular}{lccccccl}
\toprule
\textbf{Method} & \textbf{FC (gal)} & \textbf{Efficiency} & \textbf{Final SOC} & \textbf{Episodes} & \makecell[c]{\textbf{FC Reduction} \\ \textbf{vs. Conventional}} & \makecell[c]{\textbf{FC, $\eta$ vs. DP} \\ \textbf{(\% Achieved)}} & \textbf{Notes} \\
\midrule
1. Conventional & 1.95 & 41\% & -- & -- & -- & -- & Baseline reference \\

2. DP & 1.70 & 43.67\% & 83.6\% & -- & 12.8\% $\downarrow$ & 100\% & Offline optimal benchmark \\
\makecell[l]{3.Baseline DDQN \\ (No Reward Shaping)} & 0.30 & 20--25\% & 70\% & 9{,}000 & \textbf{84.6\%} $\downarrow$ & -- & very low engine power, low efficiency \\
\makecell[l]{4.DDQN \\ (Reward Shaping)} & 1.75 & 43\% & 83.6\% & 37{,}000 & 10.2\% $\downarrow$ & 97.0\% ,98.4\% &higher engine efficiency \\
\makecell[l]{5.DDQN \\ (Reward Shaping + DP Seeding)} & 1.75 & 43\% & 83.6\% & \textbf{25{,}000} & 10.2\% $\downarrow$ & 97.0\%, 98.4\% & \textbf{32.4\% faster convergence} \\
\makecell[l]{6.DDQN \\ (Reward Shaping + RB and DP Seeding)} & 1.76 & 43\% & 83.7\% & \textbf{32{,}000} & 9.7\% $\downarrow$ & 96.5\%, 98.6\% & \textbf{13.5\% faster compared to no preseeding} \\

\bottomrule
\end{tabular}%
}
\label{tab:summary_results}
\end{table*}
\section{Preseeding Replay Buffer with Expert Data}

To improve convergence speed and investigate the influence of offline data quality, the replay buffer was pre-seeded with expert demonstrations from three distinct policies: (1) Dynamic Programming (DP), and (2) a hybrid combination of dynamic programming and rule-based policy. This experimental setup aims to evaluate how different levels of policy optimality affect both convergence behavior and final policy performance.

In the first case, expert data were obtained from offline optimal DP solutions computed over various duty cycles. Approximately 25\% of the replay buffer was filled with this data before training began. Each experience was stored as a tuple in the form $(s, a, r, s')$, representing the state, action, reward, and next state. The DP trajectories were generated by minimizing cumulative fuel mass flow rate using the cost function:
\[
J = \min \left( \sum \dot{m} \, \Delta t \right)
\]

To ensure compatibility with the learning framework, the DP transitions were re-evaluated using the DDQN reward shaping function and subsequently inserted into the buffer for use during early training episodes.

  In the second case, A rule-based policy is developed where the controller uses the engine when the power required is above 110 kW and uses the battery otherwise. The resulting power split decisions were processed through the reward shaping function to align with the DDQN structure and then added to the replay buffer. Transitions from both the DP and rule-based policies were combined and shuffled to create a mixed dataset. This composite buffer was preloaded before training began to assess the impact of heterogeneous offline data on learning performance. 
  
  The replay buffer size is 200000 transitions, the batch size is 64. nearly 25\% of it is filled in the first case and nearly 50\% in the second case, starting with $\epsilon=0.5$.

\subsection{Preseeding Replay Buffer Results}

\begin{figure}[h]
    \centering
    \includegraphics[width=\linewidth]{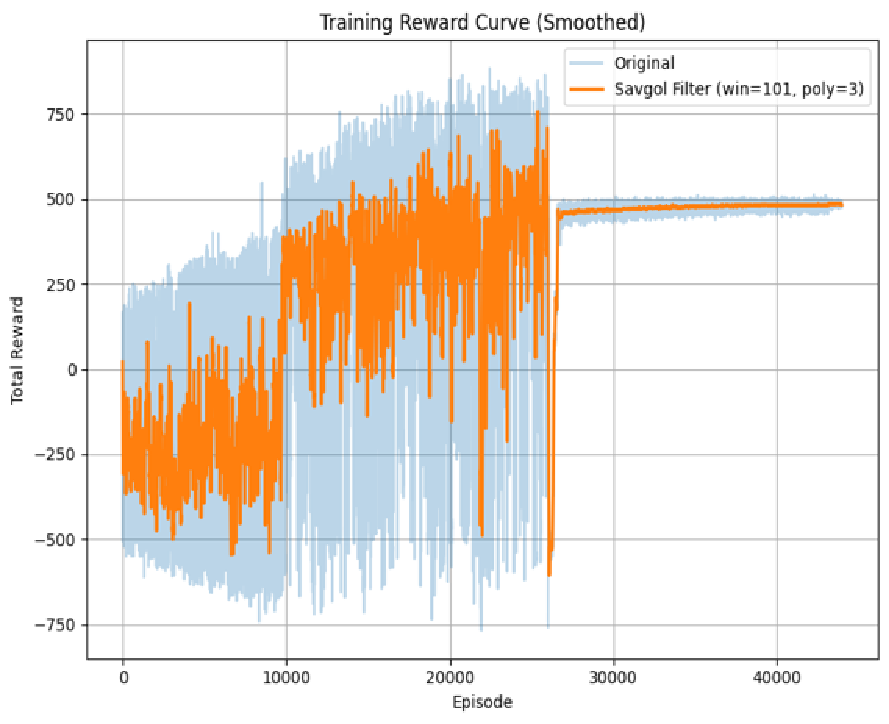}
    \caption{Reward plot for reward shaping DDQN with preseeded DP results}
    \label{fig:DDQN_DP_reward}
\end{figure}

\begin{table}[h]
\centering
\caption{Comparison of Methods Based on Fuel Consumption (FC), Efficiency, Final SOC, and Convergence}
\small
\resizebox{\columnwidth}{!}{%
\begin{tabular}{lcccc}
\toprule
\textbf{Method} & \textbf{FC (gal)} & \textbf{Efficiency} & \textbf{Final SOC} & \textbf{Episodes} \\
\midrule
\makecell[l]{1.DDQN\\(No expert data)} & 1.75 & 43\% & 83.6\% & 37{,}000 \\
\makecell[l]{2.DDQN\\(DP Data)} & 1.75 & 43\% & 83.6\% & 25{,}000 \\
\makecell[l]{3.DDQN\\(DP+RB Data)} & 1.76 & 43\% & 83.7\% & 32{,}000 \\
\bottomrule
\end{tabular}%
}
\label{seeding_results}
\end{table}

\subsubsection{ Preseeding the Replay buffer with DP Policy:} Figure~\ref{fig:DDQN_DP_reward} shows that the system converges after approximately 25{,}000 episodes, representing a 32.4\% reduction in training time compared to the case without expert seeding. As shown in Table~\ref{seeding_results}, this improvement in convergence speed does not compromise the optimality of the learned policy. Both DDQN models, with and without expert seeding, achieve the same final SOC of 83.6\%, indicating consistent energy balance across policies.

\subsubsection{Pre-seeding Replay Buffer with Combined Data:} 
\begin{figure}[h]
    \centering
    \includegraphics[width=\linewidth]{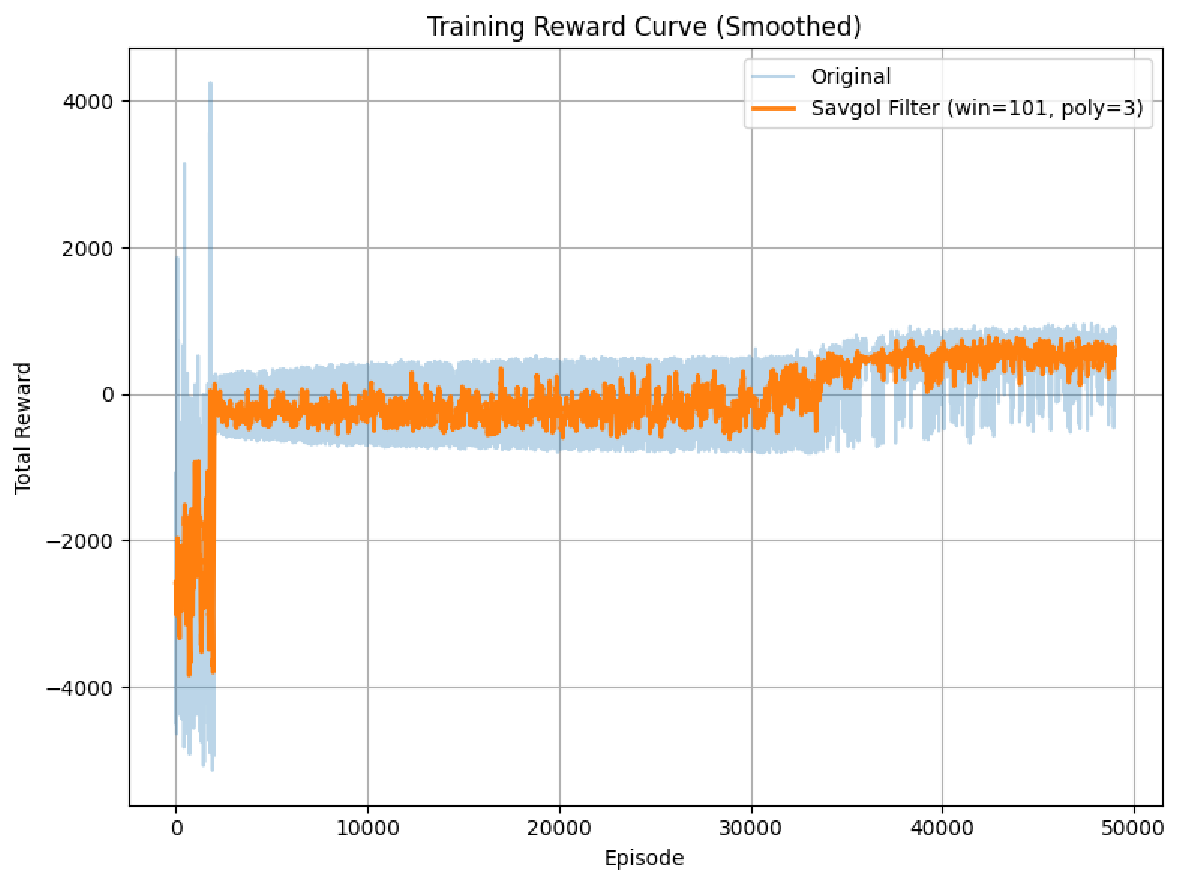}
    \caption{Reward plot for reward shaping DDQN with preseeded DP and RB results}
    \label{fig:DDQN_DP_RB_seeding_reward}
\end{figure}

Figure~\ref{fig:DDQN_DP_RB_seeding_reward} shows that the agent converges to the same policy after approximately 32{,}000 episodes when the replay buffer is pre-seeded with a combination of DP and rule-based data. This convergence speed lies between that of DP-only seeding (25{,}000 episodes) and rule-based seeding (40{,}000 episodes), indicating that mixed-quality offline data can still accelerate training while preserving policy performance. These results suggest that while expert-level data (e.g., DP) provides the fastest learning, combining it with simpler rule-based data remains an effective and practical compromise when optimal data is limited or costly to obtain.

Although the inclusion of expert data does not further reduce fuel consumption or improve efficiency, it offers a clear practical benefit by accelerating the training process. These results indicate that the quality of the data being pushed into the replay buffer affects convergence speed, but not optimality. These results confirm that initializing the replay buffer with expert-guided data effectively enhances convergence without degrading performance, validating the utility of the proposed hybrid learning framework.

Table~\ref{tab:summary_results} presents a comparative evaluation of all tested methods in terms of fuel consumption, engine efficiency, final SOC, convergence speed, and relative performance. The Dynamic Programming (DP) solution serves as the optimal benchmark, achieving the lowest fuel consumption (1.70 gal) and highest efficiency (43.67\%). The baseline DDQN without reward shaping exhibits minimal fuel usage due to excessive reliance on the battery but suffers from poor engine efficiency and low final SOC, making it impractical. Incorporating reward shaping significantly improves efficiency to 43\%, aligning closely with the DP benchmark, while maintaining a high final SOC and yielding a 10.2\% fuel reduction compared to the conventional strategy.

Notably, pre-seeding the replay buffer with DP data accelerates convergence by 32.4\% without compromising performance. When both rule-based and DP data are used for buffer initialization, convergence occurs after 32,000 episodes—faster than reward shaping alone and only slightly behind DP-only seeding—demonstrating a 13.5\% speedup. This highlights the effectiveness of combining expert and heuristic data to improve training efficiency while preserving near-optimal performance, achieving 96.5\% of DP’s fuel consumption and 98.6\% of its efficiency.

\section{Conclusion}
This paper evaluated the performance of Double Q-Learning (DQL), Deep Q-Network (DQN), and Double Deep Q-Network (DDQN) for hybrid vehicle energy management. A reward shaping function was developed to encourage operation in high-efficiency engine regions, and the impact of pre-seeding the replay buffer with expert trajectories—sourced from Dynamic Programming (DP), a rule-based policy, and their combination—was systematically investigated. The results demonstrate that tabular DQL is unsuitable for this application due to the prohibitively large state-action space. In contrast, both DQN and DDQN successfully converged. Reward shaping significantly improved engine efficiency from 25\% to 43\%, supporting more balanced fuel-electric power usage. Additionally, pre-seeding the replay buffer with DP data reduced convergence time by 32.4\% without compromising policy performance. For the evaluated duty cycle, DDQN with and without pre-seeding achieved a fuel consumption of 1.75 gallons—only marginally higher than the 1.70 gallons achieved by the DP benchmark—while maintaining the same final SOC. Combined rule-based and DP data further improved convergence by 13.5\% over unseeded DDQN, while still maintaining near-optimal fuel and efficiency performance. Overall, these results highlight the effectiveness of reward shaping and expert-guided replay buffer initialization in accelerating learning and improving policy quality. Future work will focus on validating these findings across a broader range of duty cycles and operating conditions, and on assessing the generalization and transferability of the learned policies to other vehicle platforms.

\bibliography{aaai2026.bib}



\end{document}